\documentclass[journal]{IEEEtran}

\usepackage[utf8]{inputenc}
\usepackage{cite}
\usepackage{setspace}
\usepackage{epsf}\epsfverbosetrue
\usepackage{graphics,epsfig}
\usepackage{graphicx,epsfig}
\usepackage{epsfig}
\usepackage{multirow}
\usepackage{alltt}
\usepackage{subfigure}
\usepackage{tcolorbox}
\usepackage{float}
\usepackage{url}
\usepackage{graphicx}
\usepackage{lscape}
\usepackage{verbatim} 
\usepackage{footnote} 
\usepackage{latexsym} 
\usepackage{sidecap} 
\usepackage{wrapfig}
\usepackage{eso-pic}
\usepackage{fix-cm}
\usepackage{algorithm}
\usepackage{algpseudocode}
\usepackage{color}
\usepackage{wrapfig}
\usepackage{multicol}

\usepackage{flowchart}
\usetikzlibrary{shapes,arrows,matrix,decorations.pathreplacing,shapes.geometric,positioning,calc}

\usepackage{layout}

\usepackage{amsmath}
\usepackage[
  locale = DE 
]{siunitx}

\usepackage{textcomp}
\usepackage{multirow}
\usepackage{mathtools}
\usepackage{soul} 
\usepackage{amsmath} 
\usepackage{verbatim}
\usepackage{csvsimple} 
\usepackage{array}
\usepackage{subfig}
\usepackage[normalem]{ulem}
\usepackage{longtable}
\usepackage{booktabs}
\newcolumntype{P}[1]{>{\RaggedRight\arraybackslash}p{#1}}
\newcommand{\tabitem}{\textbullet~~}

\usepackage{pifont}
\newcommand{\tick}{\ding{52}}%
\newcommand{\cross}{\ding{55}}%
\newcommand{\mult}{\ding{93}}%

\newcommand{\surcom}[1]{\textcolor{black}{#1}}
\newcommand{\tgcn}[1]{\textcolor{black}{#1}}

\newcommand{\tnrev}[1]{\textcolor{black}{#1}}
\newcommand{\revtwo}[1]{\textcolor{black}{#1}}

\begin{document}

\title{Optimizing Blockchain Based Smart Grid Auctions: A Green Revolution}

\author{Muneeb Ul Hassan, Mubashir Husain Rehmani, and Jinjun Chen
\thanks{M. Ul Hassan and J. Chen are with the Swinburne University of Technology, Hawthorn VIC 3122, Australia  (e-mail:  muneebmh1@gmail.com; jinjun.chen@gmail.com).}
\thanks{M.H. Rehmani is with the Munster Technological University (MTU), Ireland (e-mail: mshrehmani@gmail.com).}
}

\maketitle

\begin{abstract}

Traditional smart grid energy auctions cannot directly be integrated in blockchain due to its decentralized nature. \tnrev{Therefore, research works are being carried out to propose efficient decentralized auctions for energy trading. Since, blockchain is a novel paradigm which ensures trust, but it also comes up with a curse of high computation and communication complexity which eventually causes resource scarcity. Therefore, there is a need to develop and encourage development of greener and computational-friendly auctions to carry out decentralised energy trading.} In this paper, we first provide a thorough motivation of decentralized auctions over traditional auctions. Afterwards, we provide in-depth design requirements that can be taken into consideration while developing such auctions. After that, we analyse technical works that have developed blockchain based energy auctions from green perspective. Finally, we summarize the article by providing challenges and possible future research directions of blockchain based energy auction from green viewpoint.

\end{abstract}



\tikzstyle{decision} = [diamond, draw, fill=blue!50]
\tikzstyle{line} = [draw, -stealth, line width= 0.4mm]
\tikzstyle{elli}=[draw, ellipse, fill=red!50,minimum height=5mm, text width=5em, text centered]
\tikzstyle{block} = [draw, rectangle, fill=blue!20, rounded corners, minimum height= 10mm, minimum width = 20mm, text width=11em, text centered]

\tikzstyle{smallblock} = [draw, rectangle, fill=green!30, rounded corners, minimum height= 7mm, text width=5.5em, text centered]

\tikzstyle{newblock} = [draw, rectangle, fill=green!30, rounded corners, minimum height= 7mm, text width=5.5em, text centered]

\tikzstyle{leftblock} = [draw, rectangle, fill=yellow!60, rounded corners, minimum height= 10mm, text width=2.5em, text centered]


\section{Introduction}

\tnrev{With the advent of distributed energy resources (DERs), many traditional smart grid consumers are being converted into prosumers which have the capability to not only use but also generate the surplus energy for selling purposes. This advancement in buying and selling of surplus energy has led to the development of modern day markets for electricity, that are energy markets~\cite{tgcnrev01}. Majority of these energy markets use the phenomenon of auction based energy trading in order to handle multiple transactions and allocations in a smooth pattern~\cite{tgcnrev03}. Various energy auctions having different advantages have been developed by researchers to carry out energy trading, however, these auctions suffer with a common issue of network-wide trust. For instance, in order to carry out an energy auction one have to completely depend upon a centralized auctioneer, which in most of cases is centralized grid utility. In order to overcome this issue of trust in auctions, the phenomenon of blockchain came up as rescuer due to its decentralized nature.}\\
Till now, plenty of blockchain based auction approaches have been developed by researchers and certain traditional auctions such as double price, Vickrey, and first price auctions, etc, have been integrated into blockchain based energy trading, but this domain still faces certain issues and one of the major issue is the scarcity of resources. Usually, smart meters acting as blockchain nodes are computationally \tgcn{inefficient to carry out complex consensus, or sometimes, the storage capability of certain nodes is not effective enough} to store the continuously generated data on ledger. Therefore, there arises a need to develop computationally  efficient mechanisms that can be integrated with decentralized grid scenarios.

\begin{table*}[htbp]
\begin{center}
 \centering
 \footnotesize
 \captionsetup{labelsep=space}
 \captionsetup{justification=centering}
 \caption{\textsc{\\ \small{Comparison Summary of Previously Published Survey Articles with from Perspective of Contribution and Scope}. \scriptsize{\tick ~means the topic is covered, \cross ~shows the domain is not covered, and \mult ~means that domain is partially covered.}\scriptsize{Acronyms: Smart Grid Auctions (SGA), Green Smart Grid (GSG), Green Energy (GE), Blockchain Incentives (BI), and Types of Blockchain Auctions (ToBA).} }}
  \label{tab:surveytab01}
  \begin{tabular}{|P{2.7cm}|P{0.6cm}|P{0.7cm}|P{6.7cm}|P{0.6cm}|P{0.6cm}|P{0.6cm}|P{0.6cm}|P{0.6cm}|}
  \hline

&  & & &  \multicolumn{4}{c}{\centering  \bfseries ~~~~~~~~~~~Scope} &\\
\cline{5-09}
\rule{0pt}{2ex}
\centering  \bfseries Major Domain  & \centering  \bfseries Ref. & \centering  \bfseries Year & \centering  \bfseries Contribution Summary & \centering \bfseries SGA & \bfseries GSG & \centering \bfseries GE  & \centering \bfseries BI & \bfseries ToBA\\

\hline

\rule{0pt}{2ex}
\centering \textbf{Blockchain in Smart Cities} & ~\cite{survey01} & 2019 & A comprehensive review of application and potential of blockchain in smart cities. & \centering \mult & \centering \cross & \centering \mult  & \centering \tick & \mult  \\
\hline

\rule{0pt}{2ex}
\centering \textbf{Blockchain in Smart Grid} & ~\cite{survey02} & 2020 & A brief literature review about integration of blockchain in future smart grid. & \centering \mult & \centering \tick & \centering \tick  & \centering \mult & \cross  \\
\hline

\rule{0pt}{2ex}
\centering \textbf{Differential Privacy in Blockchain} & ~\cite{survey03} & 2020 & An in-depth survey of integration of differential privacy in blockchain layers and applications. & \centering \mult & \centering \cross & \centering \cross   & \centering \tick & \cross  \\
\hline

\rule{0pt}{2ex}
\centering \textbf{Smart Contract for Blockchain} & ~\cite{survey04} & 2020 & A systematic review of usage of smart contracts in blockchain technology. & \centering \cross & \centering \cross & \centering \cross & \centering \mult & \mult  \\
\hline

\rule{0pt}{2ex}
\centering \textbf{P2P Energy Trading} & ~\cite{survey05} & 2020 & An overview of P2P energy trading from perspective of physical \& virtual layer. & \centering \tick & \centering \cross & \centering \mult  & \centering \mult & \tick  \\
\hline

\rule{0pt}{2ex}
\centering \textbf{Green Auction Design for Blockchain based Smart Grid} & ~This Work & 2020 & A comprehensive analysis about design requirements and methodologies for overcoming resource scarcity in order to develop green auctions for blockchain based smart grid energy trading. & \centering \tick &  \tick &  \centering \tick & \tick & \tick \\
\hline

 \end{tabular}
  \end{center}
\end{table*}


\subsection{Scope and Contributions of Our Work}
Key contributions of our work are as follows:
\begin{itemize}
\item We discuss fundamental \tgcn{technologies and objectives that can play} the part in designing of green decentralized energy auctions.
\item \tnrev{We highlight designs requirements that can aid in development of infrastructure and algorithms for green auctions in blockchain based energy trading domain.}
\item We survey the existing technical works and provide a viewpoint that what aspects can be improved to make them more oriented towards green perspective.
\item We summarize certain challenges, research directions, and open issues for researchers working in the field of green auction design for blockchain based smart grid. 
\end{itemize} 

\subsection{Comparison with Related Survey Works}
Several surveys have been published that have highlighted various aspects of blockchain and smart grid~\cite{survey01, survey02, survey03, survey04, survey05}. For instance, authors in~\cite{survey01} provided a detailed overview about applications and potential trends regarding integration of blockchain in smart cities. Authors surveyed the effects of blockchain in smart grid, smart transportation, supply chain, and other similar domains. Similarly, another work that thoroughly surveyed the integration of blockchain in future smart grid architectures have been carried out by authors in~\cite{survey02}. Another work that highlights the privacy issues in blockchain and their countermeasures using differential privacy have been presented in~\cite{survey03}. Similarly, the possible use cases and the significance of smart contracts in blockchain scenario has been extensively reviewed by authors in~\cite{survey04}. One more work that provide \tgcn{in-depth insights} about peer-to-peer (P2P) energy trading in physical and virtual layer have been presented in~\cite{survey05}. Contrarily, our work highlight the aspect of green auction design in blockchain based energy trading.

\section{Fundamentals of Green Auctions for Blockchain based Energy Trading}
\tnrev{Game-theory and blockchain are playing a significant role in development of modern energy systems, which will be discussed in this section}. 
\subsection{Motivation of Blockchain based Energy Auctions}
Smart grid energy trading is advancing rapidly and auction mechanisms are playing a vital role in this development. \tnrev{In an energy auction, it is important to incentivize each participant in the best manner in order to maintain their motivation level~\cite{funref01}. These energy auction mechanisms ensure truthfulness, but due to central entity, they lacks a perception of trust among participants. To overcome this trust issue, decentralized blockchain technology came up as a rescuer~\cite{tgcnrev02}. In blockchain based auction mechanisms, all blockchain nodes have a copy of distributed ledger, thus one entity cannot control whole data.} This decentralized storage provides transparency and ensures that none of auction record can be altered once it gets recorded because everyone have a copy. Moreover, this record is maintained and updated via strong consensus mechanism that is is carried out among mining nodes.

\subsection{Types of Auctions in Green Smart Grid}
\tnrev{A vast amount of research have been carried out in the field of decentralized blockchain based energy trading. We categorize these blockchain based energy auction into three categories, which are discussed in this section.}

\subsubsection{Double Auction}
In double auction, buyers compete with other buyers to purchase energy from a determined slot and sellers compete with other sellers to sell energy in order to enhance their utility. \tnrev{In this way, an energy demand-supply curve is formed which is used to predict the outcome and market clearing price (MCP) of energy auction (c.f.~\cite{funref05} for basic auction terminologies). In blockchain energy auctions, the complete process of MCP calculation is carried out in a decentralized manner, so every buyer and seller can view and verify the demand-supply curve to ensure authenticity.} After finalizing MCP, the winning buyers pay the amount they bid ($P_j(S) = b_j$) for the specified slot, which is also known as buyers payment.

\subsubsection{Vickrey Auctions}
\tnrev{Vickrey auctions lie in the category of sealed-bid auctions, where buyers do not share their valuation/bids publicly and instead submit their valuations to some trusted auctioneer.} Generally, Vickrey auctions are divided into two further types named as \textit{k-th-price} auction and Vickrey-Clarke-Groves (VCG) auction. 
\paragraph{\textit{K-th-price}  Auction}
\tnrev{In this type of auction, the winner is the one who has the highest bid.} However, the final price $p_j$ depends upon the value of $k^{th}$ bid, and this value of $k$ determined before starting of auction. For instance, in Vickrey second price auction ($k$ =2), the highest bidder wins the energy slot, however, it will pay the second highest bid.

\paragraph{VCG Auction}
\tnrev{VCG auction is a game-theoretic generalized form of Vickrey auction, in this auction, buyer pays the amount of harm it causes to other buyers due to its presence~\cite{tgcnrev05}.} E.g., in a decentralized VCG auction, $n$ energy buyers submit their valuations $V = \{b_1, b_2, . . . ,~b_n\}$ for a specific energy slot ($s$) to the network for winner determination and price section. After completion of the specified time along with reception of all bids,  the VCG mechanism is used to determine the final price $P = \{P_1, P_2, . . . ,~P_n\}$ and allocation vector for the bidders. \tnrev{The price of each energy buyer is determined on the basis of harm its valuation causes to other buyers, the formula for calculation of price is given in Eq.~\ref{vcgeq}.}
\begin{equation}
\label{vcgeq}
P_{i}(B) = \underbrace{ \max_{b \in V} \sum_{k \neq j} B_k (b)}_{\substack{(A) \\ without~winner~k} }  - \underbrace{ \sum_{k \neq j} B_k (b^*)}_{\substack{(B)\\with~winner~k }}
\end{equation}
In the above equation, $b^*$ is the output of winner chosen with respect to highest bid, $k$ works as an iterative parameter which iterates through all submitted bids except for the winning bid from bidder $j$. In equation, part (A) is the accumulated sum of all the bids without participation of winner $j$, while part (B) is the accumulation of bids including the winning bid. 

\subsubsection{\tgcn{Conventional First Price Auctions}}
\tgcn{Apart from two major auction types, other auctions in smart grid energy trading can be categorised under the category of conventional/traditional first price auctions. In conventional first price auctions, highest bidder wins and pays the bidding price for the energy slot. However, the steps to carry out auction can vary. For instance, in sealed bid first price auction, all bids are hidden. Contrarily, in open-cry first price auction all bids are public at the time of auction. A discussion about integration of first price auctions in green blockchain based energy trading has been given in Section.~\ref{MiscRef}.} 

\subsection{Objectives of Energy Auctions}
Every auction mechanism has some objectives, for instance, some provide maximum social welfare, while others incentivize sellers, etc. \tnrev{However, five objectives can be defined as universal auction objectives that usually all blockchain energy auction mechanisms try to achieve, which are as follows:}

\subsubsection{Social Welfare Maximization}
In a decentralized energy auction, social welfare can be termed as sum of individual utilities of all participants, and individual utility is the amount of benefit every participant gets while participating in the auction process.  Similarly, the difference between final price and ask of an energy seller is termed as utility of prosumer or energy seller. A utility value greater than zero shows that the profit is gained by the particular participant. In this way, the summation of utilities of all participants is said to be social welfare of the network. Therefore, energy auction mechanism are designed in such a manner that they maximize the social welfare of the whole network.
\subsubsection{Individual Rationality}
An energy auction process will be individually rational when each participating buyer and seller have non-negative utility at the end of auction. It is done by developing energy auctions in such a manner that buyers and sellers are charged and paid exactly according to their bids and asks respectively. Therefore, it is important to keep the aspect of individual rationality as a significant parameter while development of energy auctions~\cite{funref08}.
\subsubsection{Equilibrium}
Equilibrium in auctions is a concept of game theory which is used to analyse the behaviour of participating buyers and sellers, this concept is further used for various statistical processes such as market analyzation, etc. Multiple equilibrium types have been developed by researchers to maximize benefit of auction such as Bayesian equilibrium, mixed-strategy equilibrium, correlated equilibrium, etc. However, the most commonly used equilibrium solution in energy based auctions in Nash equilibrium, which makes sure that none of the participant can get more benefit by changing their playing strategy unilaterally provided that the strategies of other participants remains constant. While developing an energy auction, researchers aim to achieve the most desired equilibrium type by developing incentive strategies.

\begin{figure*}[]
\centering
\includegraphics[scale=0.47]{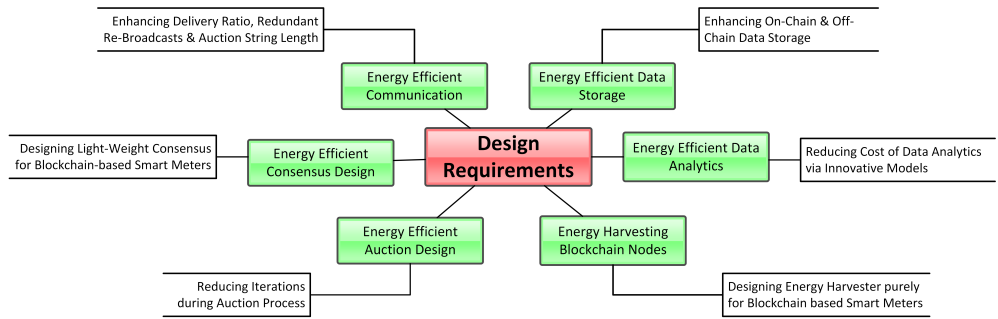}
\caption{\tnrev{Graphical Illustration of Design Requirements for Green Auctions Design in Blockchain based Smart Grid.}}
\label{Fig:Design}
\end{figure*}

\subsubsection{\tgcn{Transaction Cost Enhancement}}
\tgcn{Another objective that blockchain based energy auctions are trying to meet is to reduce to the size of auction transactions to as much degree as possible. Since blockchain is a decentralized distributed ledger and every node has a copy of the ledger, but all blockchain nodes do not have that much of liberty to store millions of records. Therefore, in order to meet the need of ever-growing auction data, research is being carried out to reduce the transaction size in order to save transaction storage cost. This aspect of transaction cost enhancement is further elaborated from perspective of energy efficient storage in Section.~\ref{storagesec}.}
\subsubsection{\tgcn{Enhancing Energy Demand and Consumption}}
\tgcn{Efficient energy utilization plays an important role during development of blockchain based energy auctions because it determines the effectiveness of any decentralized auction mechanism. Within the domain of efficient energy utilization, researchers especially focus over two aspects. \tnrev{First, from perspective of meeting energy demand of an area/suburb, and second to maintain a healthy ratio between consumption of energy with respect to its generation from renewable resources.} This objective of efficient energy usage is further divided to develop various design requirements in energy auctions, which is discussed in Section~\ref{DesignLabel}.}
\subsection{\tnrev{Choosing Optimal Auction}}
\tnrev{Since, all three auction types have their own advances and drawbacks, it is not possible to point a specific auction being best among all. Contrarily, it can be mentioned that the choice of auction basically depends upon the requirement of energy trading in smart grid. For example, if devices have less computational power and one wants to keep things simple and straightforward, then traditional first price auction is most suitable. However, if one want to allocate surplus energy while matching demand-supply curve, then double auction is the most suitable. Similarly, if one want to maximize the revenue and social welfare of buyers and sellers via game-theoretic methods then Vickrey auctions are most suitable. Overall, it will not be wrong to say that all auctions have their pros and cons, and they should be chosen depending upon requirement.}
\subsection{\tnrev{Use Case of Green Blockchain based Energy Auction}}
\tnrev{In order to demonstrate our vision of green revolution in blockchain based smart grid energy trading, we took a generalized case of a prosumer, who is interested to share his energy via blockchain based grid auction to ensure trust in transactions of network. However, if we take traditional blockchain based smart grid network, which works over proof-of-work (PoW) consensus to mine the block in the network, then this prosumer has to spend a vast amount of energy to carry out consensus. Contrarily, if one opt for a greener consensus, such as PoS or other lighter consensus variants, then the amount of energy spent is fairly less comparatively. Similarly, if the prosumer goes for traditional VCG auction, then one has to carry out a vast number of iterations before reaching to final price. Contrarily, if one uses a lighter version of VCG or any other light-weight auction, then the energy consumed during iterations can be heavily reduced. These small modifications can make a huge difference to overall environment if they are done in the right way. Therefore, the aim of this study is to motivate researchers to develop greener strategies for blockchain based smart grid energy auctions.}

\section{Design Requirements for Green Auctions in Blockchain based Smart Grid Energy Trading}\label{DesignLabel}
\subsection{Energy Efficient Blockchain Communication}
Since blockchain works over the phenomenon of decentralized distributed ledger, therefore, every node of blockchain will be connected with each other via some wireless technology. \tnrev{Therefore, every new update in the network is broadcast to all other participating nodes.} This redundant retransmissions and broadcast causes certain communication related issues such as high energy usage, channel attenuation, noise, etc. Certain communication strategies and mechanisms have been used by researchers to overcome communication cost. \tnrev{For instance, usage of bloom filters, push and announce requests, flooding, gossiping, etc~\cite{designref02}.} However, research has not been carried out from perspective of sharing auction data in the most proficient manner.\\
\indent \textit{Discussion:} Overcoming communication cost is one of the most significant challenge in blockchain systems because of redundant transmissions in case of every new update. Plenty of aspects in blockchain based auctions can be enhanced to perform energy efficient communication. Firstly, reduction of string length in auction message broadcast can play a significant role. For instance, finding the best combination of data that conveys all parametric requirements of auction can be found. Secondly, usage of green communication mechanisms which either use minimum energy or use the available spectrum in an efficient manner needs to be taken care of. For example, blockchain nodes can be converted to cognitive radio nodes in order to utilize spectrum efficiently.

\subsection{Energy Efficient Consensus for Blockchain}
Consensus mechanism is considered to be the heart of any blockchain network. However, the first blockchain consensus mechanism (PoW) required miners to solve computationally complex puzzle to mine the block. Afterwards, certain other mechanisms such as proof of stake (PoS), proof of burn (PoB), proof of Elapsed Time (PoET), etc, have been developed for different applications.\\
\indent \textit{Discussion:} Consensus mechanisms play an integral part in sustainability of blockchain network, but development of a consensus mechanism that incorporate various parametric aspects of blockchain based smart grid auctions needs to be considered. Authors in~\cite{designref05} developed an energy oriented consensus for P2P energy trading which can be considered as a first step toward green consensus design purely for energy trading. However, there is still a large room which can be explored, and researches can be carried out in plenty of directions of green consensus design. \tnrev{For instance, one direction could be to develop energy efficient consensus which is purely oriented for blockchain based auctions.} Another approach could be to develop such consensus mechanisms which work over energy generated from renewable energy resources (RERs). Overall, we believe that green consensus design can serve as a prospective direction to make blockchain based auction greener. 

\subsection{Energy Efficient Auction Design}
\tnrev{If one wants to enhance the performance of blockchain based energy trading model from green perspective then the first thing is to enhance the performance of auction mechanism in a way that it lies within energy efficient auctions.} \\
\indent \textit{Discussion:} \tnrev{As discussed in earlier sections that auction theory is a pretty old domain of statistics, however, research works are still being carried out to develop energy efficient auctions.} For instance, single unit first price auctions carried out for 500 energy slots and 500 bidders will take pretty less computational resources as compared to VCG auction. However, the game-theoretic aspects such as truthfulness, rationality, etc, of VCG auction are more dominant over first price auction. Therefore, while development of energy auctions for blockchain based smart grid, research need to consider the aspect of energy and computational consumption along with providing rationality, truthfulness, and equilibrium. 

\subsection{Energy Efficient Auction Data-Storage (On-Chain \& Off-Chain)}\label{storagesec}
Blockchain data storage has been in discussion since the advent of blockchain because data of blockchain is increasing with the addition of every new blocks in the ledger. This type of storage is required to ensure trust, however, certain nodes are not capable of storing massive amount of data. Therefore, researches are focusing over both type of storages, on-chain and off-chain.\\
\indent \textit{Discussion:} From perspective of blockchain based smart grid auctions, both of the data storing approaches can be considered as both have their pros and cons. For instance, if one store complete auction and trading data on-chain, then the nodes (especially smart meters) needs to have high storage capability to store it. Contrary to this, if the complete data is stored off-chain then one gets deprived of various useful advantages of blockchain such as trust enhancement, etc. Therefore, one need to consider both pros and cons of each strategy while development of auction. 
\subsection{Green Auction Data Analytics for Future Operations}
Data analytics is not directly linked with auction design, but it definitely has links with the stored data. As discussed in the above subsection that green methods needs to be adopted for auction data storage, similarly, the integration of green data analytic strategies while analysing auction data can play a major role in development of a complete green and cost friendly system~\cite{designref07}.\\
\indent \textit{Discussion:} Data analytics play a critical role in controlling overall cost of network because of the reason that majority of auction data is collected for the purpose future forecasting. E.g., how much energy is generated from a specific resource, or how much energy is purchased in a specific suburb, etc. Therefore, development of such data analytic mechanisms which are energy and resource efficient needs to be considered. One direction could be development of innovative architecture that utilizes energy resources in the most efficient manner, another direction could be to design such data processing frameworks which consumes minimum resources to analyse maximum data, another direction could be integration of RERs based energy in analytics, etc. \tnrev{Therefore, strategies need to be developed and action needs to be take in direction of green data analytics for blockchain based smart grid auctions.}

\subsection{Energy Harvesting Nodes for Blockchain Auction Network}
Harvesting energy from various environmental factors to power electronic devices is one of the most promising strategy to eliminate battery dependency. In order to so, more than 5000 different research works have been carried out till now~\cite{designref08}. However, harvesting energy for blockchain nodes is still a new domain and very minimal work has been carried out in this direction. Although, it is a new domain, but this concept of energy harvesting blockchain nodes have strong roots and can help in a much deeper manner towards development of green blockchain based smart grid auctions.\\
\indent \textit{Discussion:} In order to incorporate the concept of energy harvesting in blockchain based energy auctions, firstly, it is important to analyse and identify potential deployment places. For instance, one potential framework could be to integrate the concept of energy harvesting with blockchain based smart metering nodes, so that these nodes could power themselves at the time of auction and consensus process. Another possible direction could be to integrate energy harvesting only with controlling/mining nodes which are responsible to carry out extensive computational tasks. The second important aspect while integrating energy harvesting phenomenon in blockchain based auction is to figure out optimal material and manufacturing details for harvesting devices. Similarly, the aspect of cost can also not be ignored while designing energy harvesting devices for blockchain based auction systems. However, it is worthwhile to mention that the aspect of integration of energy harvesting with blockchain based energy auctions could play a significant role if its carried out in a proper direction.


\begin{table*}[htbp]
\begin{center}
 \centering
 \scriptsize
 \captionsetup{labelsep=space}
 \captionsetup{justification=centering}
 \caption{\textsc{\\ \small{ Technical Works in Blockchain based Green Smart Grid Auctions. \\ Acronyms: Social Welfare (SW), Individual Rationality (IR), Equilibrium (EQ),  Transaction Cost (TC), Energy Consumption (EC), Energy Demand (ED), Smart Contract (SC), Not Specified (N/S). }}}
  \label{tab:techtab01}
  
  \begin{tabular}{|P{0.95cm}|P{1.3cm}|P{0.4cm}|P{3cm}|P{2.4cm}|P{1.2cm}|P{0.7cm}|P{0.3cm}|P{0.3cm}|P{0.3cm}|P{0.3cm}|P{0.3cm}|P{0.3cm}|P{0.3cm}|}
  \hline

&  & & &  & & & \multicolumn{6}{c}{\centering  \bfseries ~~~~\tgcn{Auction Objectives Met}} &\\
\cline{8-14}
\rule{0pt}{2ex}
\centering  \bfseries \surcom{Auction Type}  & \centering  \bfseries \surcom{Addressed Green Issue} & \centering \bfseries Ref. & \centering  \bfseries Major Contribution & \centering  \bfseries \surcom{Type of Design Requirements Met}  & \centering  \bfseries Blockchain Type & \centering \bfseries Cons- \newline ensus & \bfseries SW & \bfseries \surcom{IR} & \bfseries \surcom{EQ} & \centering \bfseries TC & \centering \bfseries EC & \centering \bfseries ED & \bfseries SC\\

\hline
\multirow{18}{*}{\parbox{2cm}{\centering \textbf{}}}

 & \surcom{Efficient Auction Design} & ~\cite{techref11} & Localized P2P energy trading for grid connected EVs. & \tabitem Auction Iteration Reduction \tabitem Reduction in Energy Consumption & Consortium & PoW  & \tick & \surcom{\tick} & \surcom{\tick} &  &  &  &  \\
\cline{3-14}

 &  & ~\cite{techref04} & Quick converging P2P auction for smart grid. & \tabitem Auction Convergence \tabitem Energy Loss \tabitem Overhead & Public & PoW  &  \tick & \surcom{\tick} & \surcom{\tick} & \tick &  & \tick & \tick  \\
\cline{2-14}



 &  & ~\cite{techref03} & Multi-tier energy auctions for distribution grids. & \tabitem Scalability \tabitem Computational Cost & Public & Business Driven  & & \surcom{\tick} & \surcom{\tick} &  & \tick &  & \tick  \\
\cline{3-14}

& \surcom{\centering Communica- \newline tion \& Computation Enhancement} & ~\cite{techref34} & Field implementation of energy market in Switzerland.  & \tabitem Communication Cost Reduction via Tendermint & Private (permissioned) & Tend- \newline ermint  & & \surcom{\tick} & \surcom{\tick} &  &  &  & \tick  \\
\cline{3-14}

 &  & ~\cite{techref32} & Proposed three approaches for energy trading. & \tabitem Computation Efficiency & Public & PoA  & & \surcom{\tick} &   &  &  & \tick & \tick  \\
\cline{2-14}


 & \surcom{Efficient Tx \& Data Storage} & ~\cite{techref12} & Multi \& internal micro grid energy trading. & \tabitem Distributed Data Storage \tabitem Reduction in Transaction Volume  & Public & N/S  & \tick & \surcom{\tick} &  & \tick & \tick &  & \tick  \\
\cline{3-14}

\textbf{Double Auctions}  &  & ~\cite{techref10} & First price, time first based double auction for microgrid energy trading. & \tabitem Improving Transaction Efficiency  & Consortium & PBFT  &  \tick & \surcom{\tick} &   &  & \tick  &  &   \\
\cline{2-14}


 &  & ~\cite{techref23} & Power flow \& multilateral energy trading via Ethereum blockchain. & \tabitem Detecting Overlimit Power-Flow to Manage Energy & Private & PoW  & & \surcom{\tick} &  & \tick &  &  & \tick  \\
\cline{3-14}

&  & ~\cite{techref14} & Decentralized load balancing for P2P energy trading. & \tabitem P2P Energy Cost Reduction & Public & N/S  & \tick & \surcom{\tick} & \surcom{\tick} &  & \tick &  &  \\
\cline{3-14}

 &  & ~\cite{techref27} & Decentralized load balancing via energy markets. & \tabitem Managing \& Trading Energy Locally & Public & Tend- \newline ermint  & \tick & \surcom{\tick} & \surcom{\tick} &  & \tick  &  & \tick  \\
\cline{3-14}

& & ~\cite{techref24} & Charging power quota based energy trading. &\tabitem Charging Stations Settlement via Quota & Private & N/S  & \tick & \surcom{\tick} &  &  & \tick &  & \tick  \\
\cline{3-14}

&  & ~\cite{techref08} & Selfish \& helpful bidding strategy for residential DERs. & \tabitem Peak Load Reduction  & PBFT & PBFT  & \tick & \surcom{\tick} &  &  & \tick & \tick & \tick  \\
\cline{3-14}

 & \surcom{Enhancing Energy Cost} & ~\cite{techref26} & Grid influenced P2P energy trading via game theory. & Reduction in Peak Demand & Public & N/S  & \tick & \surcom{\tick} & \surcom{\tick} &  & \tick  & \tick & \tick  \\
\cline{3-14}

 &  & ~\cite{techref33} & Energy exchange and settlement procedures for energy trading. & \tabitem Balancing Energy Consumption \& Production Ratio& Public & PoS  & \tick & \surcom{\tick} &   & \tick  & \tick  & \tick  & \tick  \\
\cline{3-14}

&  & ~\cite{techref35} & Decentralized local energy trading to enhance sustainability. & \tabitem Enhancing Local Trade to Prevent Energy Losses& Public & PoI  & \tick & \surcom{\tick} &  &  &  &  & \tick  \\
\cline{3-14}

&  & ~\cite{techref16} & Hybrid P2P and P2G energy trading. &  \tabitem Cost Reduction via Hybrid Energy Trading & Public & N/S  & \tick & \surcom{\tick} & \surcom{\tick} &  &  &  & \tick  \\
\cline{3-14}

 &  & ~\cite{techref02} & Reward enhancing transactive energy auctions. & \tabitem Local Energy Trading Enhancement  & Public & PoW  & \tick & \surcom{\tick} & \surcom{\tick} &  &  &  & \tick  \\
\cline{3-14}

\hline

\hline
\multirow{4}{*}{\parbox{2cm}{\centering \textbf{}}}

\textbf{Vickrey Auctions} & \surcom{Enhancing Energy Cost} & ~\cite{techref01} & Differentially private decentralized microgrid auction. & \tabitem Energy Reduction & Consortium & PoW  & \tick & \surcom{\tick} &   &  &  &  &   \\
\cline{3-14}
 &  & ~\cite{techref21} & Quasi-ideal P2P transactive energy trading. & \tabitem Managing Market Surplus Energy & Public & PoC  & \tick & \surcom{\tick} &   &  &  & \tick & \tick  \\

\hline

\multirow{5}{*}{\parbox{2cm}{\centering \textbf{}}}
 &  & ~\cite{techref05} & Multi-microgrid based energy trading. & \tabitem Efficient Architecture Design & Public & N/S  & \tick & \surcom{\tick} &  &  &  &  &   \\
\cline{3-14}

& \surcom{Enhancing Energy Cost} & ~\cite{techref06} & Transactive energy exchange for EI based blockchain. & \tabitem Energy Management via EI Concept & Private (permissioned) & PBFT   & \tick & \surcom{\tick} &  &  &  & \tick & \tick  \\
\cline{3-14}

&  & ~\cite{techref31} & Practical implementation of decentralized energy trading. & \tabitem Cost Reduction & Public & N/S  &  & \surcom{\tick} &   &  &  &  & \tick  \\
\cline{2-14}


\centering \textbf{Conven- \newline tional First} & \surcom{Efficient Tx \& Data Storage} & ~\cite{techref15} & Game-theoretic framework for V2G network. & \tabitem Tx Throughput \tabitem Scalability & Public & IoTA  & \tick & \surcom{\tick} & \surcom{\tick}  &  & \tick & \tick & \tick  \\
\cline{3-14}

\centering \textbf{Price Auctions} &  & ~\cite{techref17} & Sealed auction transactions for V2G network. & \tabitem Transaction Matching \& Convergence & Public & PoA  & \tick & \surcom{\tick} &  &  &  & \tick & \tick  \\
\cline{3-14}

 &  & ~\cite{techref19} & Blockchain + IFPS storage for energy trading. & \tabitem Enhancing Data Storage  & Public & Multiple  & \tick & \surcom{\tick} &   & \tick & \tick &  & \tick  \\
\cline{3-14}


&  & ~\cite{techref30} & DSM enhancing grid optimal auction. & \tabitem Scalability Enhancement & Public & N/S  & \tick & \surcom{\tick} & \surcom{\tick}  &  & \tick &  &   \\
\cline{2-14}


 & \surcom{Efficient Auction Design} & ~\cite{techref17} & Sealed auction transactions for V2G network. & \tabitem Transaction Matching \& Convergence & Public & PoA  & \tick & \surcom{\tick} &  &  &  & \tick & \tick  \\

\hline

 \end{tabular}
  \end{center}
\end{table*}

\section{Blockchain based Auction Approaches for Green Smart Grid} \label{BlockAuctions}
Various technical works integrating different types of auctions strategies in blockchain based energy trading have been carried out by researchers till now. On a broader scale, these works can be categorised in to three major categories: 1) double auctions 2) Vickrey Auctions 3) conventional first price auctions. \surcom{We discuss a comprehensive overview of these technical works from perspective of green design requirements in this section.}~ \tnrev{A detailed table highlighting the contribution and addressed design requirements alongside other technical parameters have been presented in Table.}~\ref{tab:techtab01}.

\subsection{Double Auction} \label{DoubleRef}
A double auction mechanism is said to be incentive-compatible if by acting on the preferred bids and asks, each buyer and seller gets their best outcome. Therefore, while developing double auction based energy trading mechanisms, researchers tend to integrate the functionality of incentive compatibility in their mechanisms. 
\subsubsection{\surcom{Efficient Auction Mechanism Design}}
\surcom{Enhancing efficiency of double auction by reducing the convergence time is a critical step towards designing of green double auctions for decentralized smart grids. Till now, two works have been carried out by researchers that focused over reducing the iteration of energy auctions in order to converge them quickly as compared to traditional double auction. One such work in the direction of efficient auction design has been carried out by Kang~\textit{et al.} in~\cite{techref11}, where authors work over enabling P2P energy trading for grid connected EVs using efficient double auction mechanism. Authors used the phenomenon of flag and trigger to solve the complex iterative problem of double auction in an optimal way. The proposed mechanism uses less iteration to converge, which in turn save excessive usage of energy during iteration process. Furthermore, the work also used consortium blockchain to overcome computational scarcity issue for the nodes which do not have significant computational resources. Another work discussing quick converging double auction have been presented by authors in~\cite{techref04}. Authors designed a mechanism via which majority of energy demands can be met in short number time, which in turn reduces the converging time for double auction. Moreover, authors also enhanced energy losses and computation overhead by motivating peers to trade energy locally.}

\subsubsection{\surcom{Communication \& Computation Cost Reduction}}
\surcom{Providing efficient communication and computation has always been an important objective while development of decentralized energy auctions. \tnrev{In order to do so, certain research works has been carried out,} one such work to provide energy efficient computation for multi-tier auction distribution grids  have been carried out by authors in~\cite{techref03}. Authors used a business driven consensus mechanism to enhance computational factor along with providing a more scalable blockchain model for secure energy auctions. Similarly,} from perspective of field implementation, a detailed study on blockchain based energy market of Switzerland has been presented by authors in~\cite{techref34}. The work discussed reduction of communication and computation cost by using Tendermint consensus instead of traditional consensus protocols.
\surcom{Another similar work that targets to reduce computational scarcity in blockchain based auctions have been carried out by authors in~\cite{techref32}. The work proposed three energy trading approaches for blockchain based smart grid and compared them on the basis of cost and effectiveness. Similarly, the authors also provided the comparison of these approaches from perspective of clearing price and clearing quantity.}

\subsubsection{\surcom{Transaction \& Data Storage Efficiency}}
\surcom{Storing and managing auction data in an efficient manner is one major challenge that blockchain based energy trading nodes are facing right now because of the distributed nature of blockchain. Since blockchain works over the phenomenon of distributed ledger, therefore, every new transaction is stored over this ledger. Considering the resource capacity of participating energy nodes, it is important to design such auction mechanisms which utilize minimum space while storing transaction data. On such work discussing} an important aspect of dealing with energy transaction from perspective of both internal and multi-microgrids have been addressed by Zhao~\textit{et al.} in~\cite{techref12}. Authors enhanced distributed data storage along with reduction in transaction volume in order to provide more sustainable blockchain network from the point of view of our design requirement of efficient data storage. \surcom{Another work to improve transaction efficiency by using first price and time-based double auction for blockchain based energy trading have been carried out authors in~\cite{techref10}. Authors evaluated the phenomenon of enhanced of transaction efficiency at difference points and claimed that the proposed mechanism provides energy efficiency alongside providing transaction storage efficiency. Furthermore, the work also used consortium blockchain PBFT consensus to reduce computational consumption.}

\subsubsection{\surcom{Enhancing Energy Cost}}
Among all design requirements, one of the most critical design requirement while developing of double sided energy auctions is to reduce energy consumption as much as possible. Researchers are doing this via various approaches, some worked over motivating users to trade energy locally, while others proposed hybrid mechanisms for this trade. In this section, we summarize all these from perspective of their particular contribution to reduce energy usage at grid or user level in blockchain based energy trading auctions. 
First work that evaluated power flow and multilateral energy trading via Ethereum have been carried out by Jin~\textit{et al.} in~\cite{techref23}. One of the major contribution of authors in this work from green perspective is detection of overlimit power flow to reduce energy losses. Similar to this, two works focusing over decentralized load balancing via energy auction have been carried out by authors in~\cite{techref14, techref27}. In~\cite{techref14}, authors reduced P2P energy trading cost by proposing an auction mechanism which works over three layered blockchain architecture including application, virtual, and physical layer. Similarly, in~\cite{techref27}, authors proposed an approach to balance renewable generation locally via novel auction market design.\\
A unique work from perspective of power quota based energy auction over blockchain have been presented by Ping~\textit{et al.} in~\cite{techref24}. Furthermore, authors worked over development of energy efficient mechanism for charging station settlement. A work that analyses the behaviours of selfish and helpful buyers on residential RER auction have been presented in~\cite{techref08}. Authors developed efficient bidding based MCP management system which enhances DSM by reducing peak load from residential houses. Similarly, an extensive theoretical contribution utilizing the benefits of Stackelberg game in blockchain based double auction have been presented by authors in~\cite{techref26}. Authors proposed a grid-influenced game-theoretic approach to reduce energy usage in peak hours which in turn provides efficient DSM. Similarly, a works on energy exchange and settlement procedures have been carried out by  Han~\textit{et al.} in~\cite{techref33}. Authors proposed a blockchain based trading architecture that works over execution of smart contract in order to balance energy consumption and production.\\
\surcom{A similar work focusing over decentralized energy trading to enhance sustainability of smart grid have been carried out by authors in\cite{techref35}. Authors motivated local prosumers and buyers to carry out local trade in order to prevent surplus energy losses. Another} work focusing over hybrid P2P and P2G energy trading blockchain based double auction have been presented by researchers in ~\cite{techref16}. In this work, authors focused over cost reduction by motivating the trend of hybrid energy trading rather than just P2P energy trading. Apart from traditional double auction, a work focusing over Bandit learning based energy trading have been presented by researchers in~\cite{techref02}. The work focused over using Bandit learning based double auction to enhance reward in transactive energy auctions in order to motivate maximum sellers to trade energy to the decentralized market.

\subsection{Vickrey Auctions} \label{VickreyRef}
\tnrev{In blockchain based green energy auctions, Vickrey auctions have been applied in two technical works, and in both of them they applied the advanced VCG auction model.} In VCG auction, highest bidder wins but pays the harm its presence have caused to other participating bidders. A detailed discussion about theoretical aspect of VCG auction is out of scope of this article, interested readers are suggested to go study the work in~\cite{techref21}. Moreover, both of the works applying VCG auction in blockchain based energy trading focused over reduction of energy cost and consumption to enhance the green effect in trading.\\

\subsubsection{\surcom{Enhancing Energy Cost}}
The first work integrating differential privacy, VCG auction, and consortium blockchain for decentralized energy auctions have been carried out by authors in~\cite{techref01}. The given work enhanced buyers and sellers utility along with providing privacy preservation via differential privacy guarantee. In order to preserve energy and communication cost of all participating nodes, authors proposed the usage of consortium blockchain instead of public blockchain. By using this blockchain network, only the selected participants will be able to take part in consensus, which in turn save the computation cost for casual smart metering nodes. Another work over VCG auction based energy trading have been carried out by authors in~\cite{techref21}. \surcom{The authors worked over quasi-ideal P2P transaction energy trading mechanism in order to manage market surplus energy via VCG auction mechanism. Authors proposed a mechanism to handle four types of energy trading mechanisms in parallel, via which users can trade in a parallel manner according to different energy requirements.}

\subsection{\surcom{Conventional First Price Auctions} \label{MiscRef}}
Apart from two major auctions, certain works used various other auctions approaches such as ascending price auction, message broadcasting base auction, Ausubel clinching auction, etc. \surcom{It is important to note that all of these works chose highest bidder as a winner. Based upon this understanding, we named the section as 'Conventional First Price Auctions'. }

\subsubsection{\surcom{Enhancing Energy Cost}}
In first price auction mechanisms, one of the major focus of researchers is to enhance energy cost in order to make sure that the proposed mechanism is suitable for energy constrained blockchain nodes. Plenty of works have been carried out to overcome this energy scarcity, one such work functioning on unified weight clearing to select hammer price for energy slots have been proposed by Li~\textit{et al.} in~\cite{techref05}. Authors designed an energy efficient architecture for smooth energy trading in multi-microgrid scenario. Similarly, another work that provides a thorough comparison between three auction strategies on the basis of tokens, supply-demand regulation, and energy consumption has been presented by authors in~\cite{techref06}. The work further used the concept of energy Internet with transactive grid to trade energy in local markets in order to prevent excessive energy losses. Similarly, another work that used traditional first price auction along with providing the practical implementation of decentralized energy trading have been presented in~\cite{techref31}. Authors presented a case study of practical implementation that ensured reduction of energy cost as compared to traditional blockchain design. 

\subsubsection{\surcom{Transaction \& Data Storage Efficiency}}
Another green design requirement that have been addressed in first price energy auction works is enhancement of transaction size and data storage. A very detailed work from perspective of enhancing Tx throughout and scalability of blockchain based game-theoretic ascending price auction have been presented by Hassija~\textit{et al.} in~\cite{techref15}. Authors first provided a thorough architecture for grid-connected EVs energy trading, and then evaluated the proposed model by showing that the model enhanced transaction throughput and blockchain scalability. One significant work that integrated IPFS storage with blockchain of manual E-auction based energy trading have been presented by authors in~\cite{techref19}. The work developed a novel system model to show the interoperability of IFPS storage with blockchain, and afterwards authors provided simulation based experiments to show that the proposed model reduces latency, cost, and transaction time as compared to traditional approaches. Similarly, a work that uses Ausubel’s clinching auction based energy trading to enhance DSM of smart grid have been carried out by authors in~\cite{techref30}. The work ensured that the proposed mechanism enhances scalability in order to run all blockchain operations in an efficient manner.

\subsubsection{\surcom{Efficient Auction Mechanism Design}}
\surcom{Developing efficient auction mechanisms, which can be replaced with traditional first price auctions is an important step towards green energy auctions for blockchain. One such mechanism that works over the phenomenon of reverse auction to increase auction design efficiency have been presented by authors in~\cite{techref17}. }This work also analysed EV auction based power trading mechanism for grid connected EVs. The proposed work enhanced transaction matching and quick convergence in order to utilize energy in the most efficient manner.

\subsection{Summary and Lessons Learnt}
Integrating decentralized blockchain with smart grid auctions have paved path for future grid networks in which users can trade their energy with full trust. However, integration of green aspects in blockchain based energy auction scenarios still require further research. For instance, the six design requirements that we discussed in Section~\ref{DesignLabel} have not been fully addressed in the works and only few works partially address these requirements. \tnrev{However, it is worthy to mention that the works involving double auction are more inclined towards meeting multiple design requirements. Similarly, from perspective of Vickrey auctions, only the aspect of energy cost reduction can be seen in the literature. Moving towards conventional first price auctions, certain works tried to enhance energy consumption and transaction cost.} \\
\surcom{To conclude, overall, a trend of enhancing energy cost can be seen in among all three auction categories, however, the category of conventional first price auction dominates in designing of efficient transaction and data storage models. Therefore, if one is interested to design an application in which nodes have limited memory and can handle less data volume, then works in the category of first price auction can play their role. Contrarily, if one’s application has nodes with good storage power and are more energy constrained, then works from double and VCG auction can play their role. }

\section{Challenges and Future Research Directions}

\subsection{Future Directions of Blockchain-based Green Energy Auctions}
\subsubsection{Energy Harvesting}
Traditional energy harvesting devices use kinetic, solar, RF, or thermal energy as a source~\cite{newref02}. \revtwo{These energy harvesting devices are being used by IoT nodes to generate energy for communication and transmission~\cite{newref03}. Contrary to IoT nodes, blockchain nodes require large energy to carry out communication especially at the time of consensus. This is because in blockchain consensus, a block goes through a lot of phases such as broadcast, verification, acknowledgement, approval, etc. In all these phases, block needs to be sent to all nodes via communication medium, which incur a lot of energy . Therefore, small energy harvesting devices are not capable to provide energy for this communication. Therefore, design and development of such energy harvesting devices which are purely developed for the purpose of blockchain based energy grid needs attention.}
\subsubsection{Integration of AI in Auctions}
Machine/deep deep learning is also being used to predict optimal revenue of an auction mechanism having fixed budget~\cite{newref04}. Since, carrying out machine/deep learning is a computationally complex task and at certain times one need computers with high processing power in order to process large amount of data. Therefore, if one want to apply machine/deep learning with blockchain based auction, then he has to traverse through all blocks in order to get required information~\cite{newref05}. \revtwo{Therefore, there is a need to design such machine/deep learning based prediction and analysis models which use minimal resources while fetching and training data from blockchain decentralized ledger. }
\subsubsection{Privacy Preservation in Green Blockchain based Energy Auctions}
Blockchain based energy networks also comes up with various privacy issues due to their transparent nature. The data of transactions and trading is publicly disseminated on ledger in order to enhance trust, but this also raises various privacy concern~\cite{myref02}. \revtwo{For instance, this data can be used by adversaries to carry out some malicious activities. \surcom{Therefore, the works which consider development/integration of green aspect in blockchain based energy trading do also needs to consider overcoming privacy issues such as transactional and consensus privacy in order to reduce any prospective risk.} Therefore, this direction of private auctions require further exploration, as very minimal work has been carried out yet. }
\subsection{Futuristic Applications of Blockchain-based Green Energy Auctions}
\subsubsection{\revtwo{Cognitive Radio based Green Communication}}
\revtwo{Excessive communication cost is one of the major hurdle in development of green networks~\cite{newref08}. To overcome this, integration of dynamic spectrum access via cognitive radio technology in blockchain based auction networks can play a vital role.} \tnrev{Research can be carried out for this integration, e.g., blockchain nodes can serve as cognitive radio nodes and can take advantage from various functionalities of cognitive radio such as using unlicensed spectrum at time of inactivity or using multiple channels to carry out auction communication, etc.} However, the most significant integration will be to merge the functionality of cognitive radio during the decentralized auction consensus because communication overhead during consensus constitutes a major part of blockchain. 

\subsubsection{\revtwo{Integrating Electric Vehicles with Blockchain based Energy Trading from Green Perspective}}
\revtwo{Apart from traditional prosumer scenarios, the direction of energy trading via electric vehicles (EVs) is also flourishing~\cite{newref09}. Alongside designing basic auction and trading models for blockchain based EV trading, it is also important to develop such models which integrate green aspect. E.g., design of efficient communication mechanism for blockchain based EVs could be one possible application. Similarly, designing an EV oriented consensus could help in development of greener energy trading via EVs. Considering this discussion, we believe that blockchain based EVs can serve as a pathway towards development of greener energy auctions via blockchain.}

\section{Conclusion}
In this article, we highlighted the need of green aspect in blockchain based energy auctions, then we highlighted certain design requirements which needs to be considered while development of such auctions. Afterwards, we analysed various technical works that has been carried out in this domain. Finally, we demonstrated challenges and possible future research directions for green integration in blockchain based smart grid auctions.

\bibliographystyle{IEEEtran}

\begin{IEEEbiography}[{\includegraphics[width=1in,height=1.25in,clip,keepaspectratio]{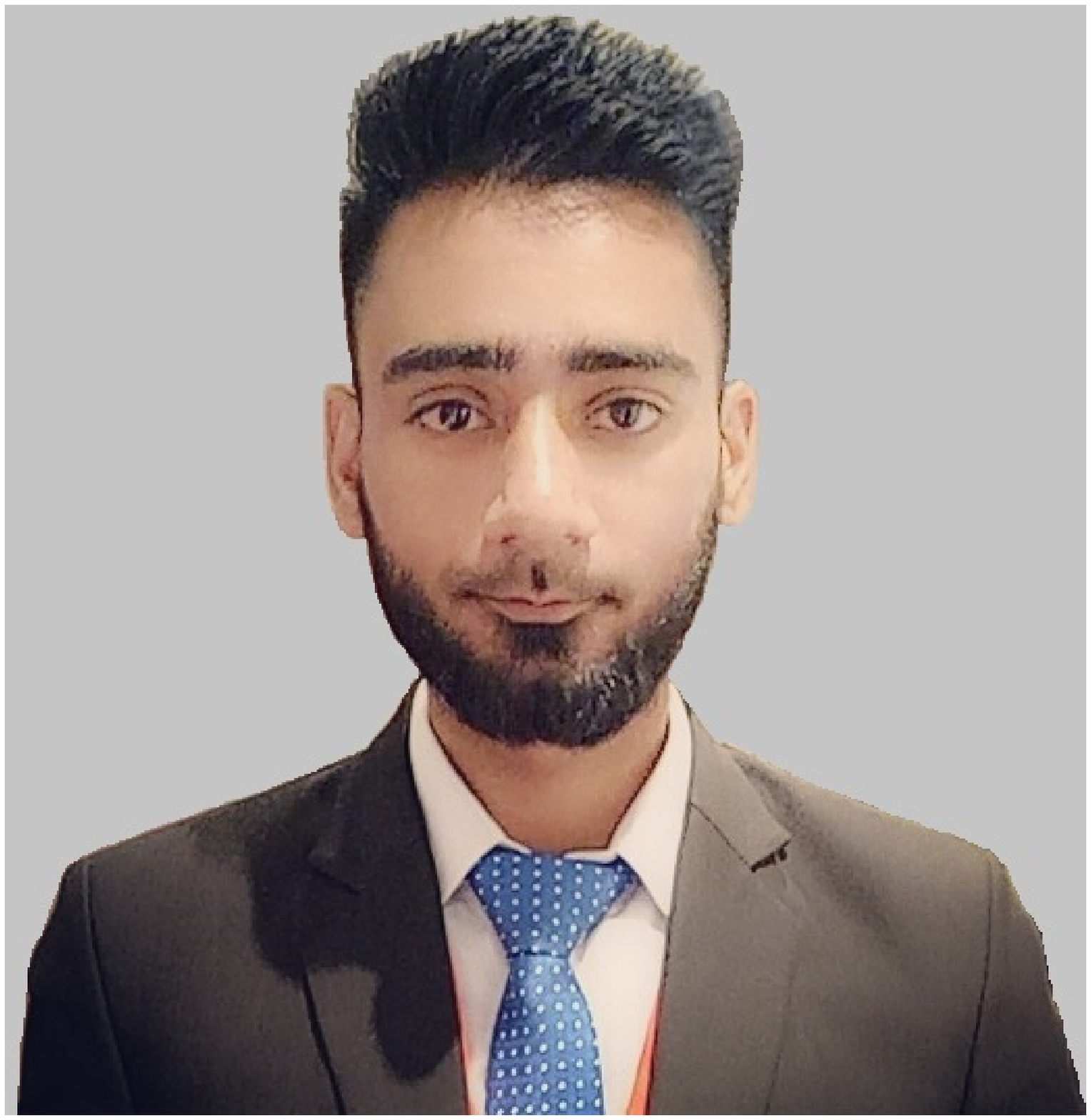}}]{Muneeb Ul Hassan} received his Bachelor degree in Electrical Engineering from COMSATS Institute of Information Technology, Wah Cantt, Pakistan, in 2017. He received Gold Medal in Bachelor degree for being topper of Electrical Engineering Department. Currently, he is pursuing the Ph.D. degree from Swinburne University of Technology, Hawthorn VIC 3122, Australia.

\end{IEEEbiography}

\begin{IEEEbiography}[{\includegraphics[width=1in,height=1.25in,clip,keepaspectratio]{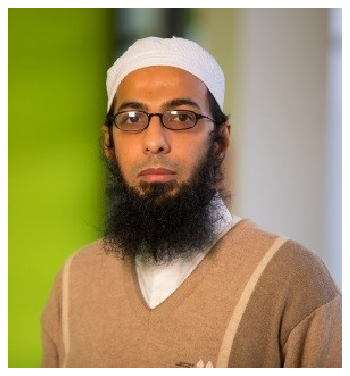}}]{Mubashir Husain Rehmani (M’14-SM’15)} received the M.S. degree from the University of Paris XI, Paris, France, in 2008, and the Ph.D. degree from the University Pierre and Marie Curie, Paris, in 2011. He is currently working as Assistant Lecturer at Munster Technological University (MTU), Ireland. He has been selected for inclusion on the annual Highly Cited Researchers™ 2020 list from Clarivate. 

\end{IEEEbiography}

\begin{IEEEbiography}[{\includegraphics[width=1in,height=1.25in,clip,keepaspectratio]{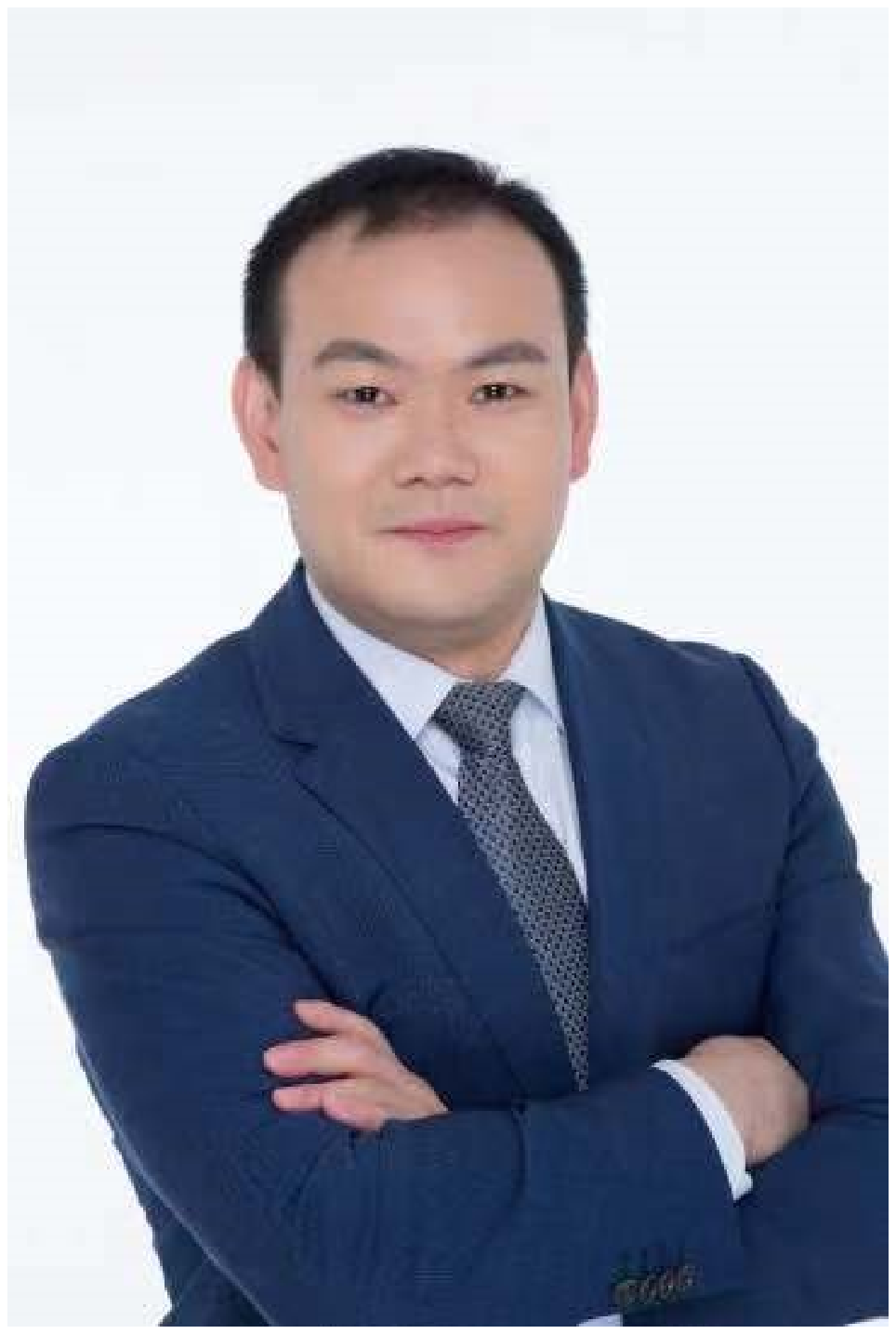}}]{Dr. Jinjun Chen} is a Professor from Swinburne University of Technology, Australia. He is Deputy Director of Swinburne Data Science Research Institute. He holds a PhD in Information Technology from Swinburne University of Technology, Australia. His research results have been published in more than 160 papers in international journals and conferences, including various IEEE/ACM Transactions. 
\end{IEEEbiography}

\end{document}